# Clocked Magnetostriction-Assisted Spintronic Device Design and Simulation

Rouhollah Mousavi Iraei, Nickvash Kani, Sourav Dutta, Dmitri E. Nikonov, Sasikanth Manipatruni,
Ian A. Young, John T. Heron, and Azad Naeemi

*Abstract*— We propose a heterostructure device comprised of magnets and piezoelectrics that significantly improves the delay and the energy dissipation of an all-spin logic (ASL) device. This paper studies and models the physics of the device, illustrates its operation, and benchmarks its performance using SPICE simulations. We show that the proposed device maintains low-voltage operation, non-reciprocity, non-volatility, cascadability, and thermal reliability of the original ASL device. Moreover, by utilizing the deterministic switching of a magnet from the saddle point of the energy profile, the device is more efficient in terms of energy and delay and is robust to thermal fluctuations. The results of simulations show that compared to ASL devices, the proposed device achieves 21× shorter delay and 27× lower energy dissipation per bit for a 32-bit arithmetic-logic unit (ALU).

## I. INTRODUCTION

OVER the past two decades, spintronic devices have been pursued to augment CMOS technology thanks to their potential advantages in terms of non-volatility, scalability, low voltage operation, and new and enhanced functionalities [1] – [3]. The proposed spin-based logic devices include the all spin logic (ASL) [4], the composite-input magnetoelectric-based logic technology (COMET) [5], the domain-wall magnetic logic (mLogic) [6], magnetoelectric spin-orbit (MESO) device [7], and the magneto-electric magnetic tunnel junction (MEMTJ) [8]. Thanks to its metallic nature and its potential back end of the line (BELO) compatibility, the ASL device has been widely studied for various applications such as CMOS/spintronic signal transduction [9], interconnection [10] - [13], image recognition [14], neural networks [15] - [16], and Boolean logic circuits [17]. However, to further improve the performance of the device, two limiting factors must be addressed. First, the ASL proposal is based on a non-local spin valve (NLSV), in which a pure spin current applies a spin-transfer torque (STT) on a free magnet and flips its orientation. Most of the spin current, however, is shunted to ground and wasted. Moreover, the experimental evidence for the operation of the NLSV is limited to only one report [18]. Second, the reliable 180⁰ switching of a magnet using STT is known to be quite slow (~ few nanoseconds) and requires large current densities [19]. Passing large currents through select transistors for such large time periods results in prohibitively large energy per binary-switching operation. As a result, even when supply clocking is used [17], a 32-bit arithmetic-logic unit (ALU) based on ASL is projected to dissipate more than four orders of magnitude more energy compared to its CMOS counterpart [20].

Recent theoretical predictions of reliable magnetic reversal demonstrate that magnets initialized at their saddle point of energy profile require far less switching energy [21]. Moreover, the energy-efficient 90⁰ rotation of magnets using magnetostriction is demonstrated [22]. Informed by the recent developments, this paper proposes a novel spintronic logic device [23] comprising of a piezoelectric-magnetic heterostructure that employs both the voltage-controlled, strain-mediated, magnetoelectric switching and the STT switching from the saddle point of energy. In contrast to ASL, the STT in the proposed logic is created by the conventional spin valve (CSV) structure for which there are many experimental demonstrations [24] – [25]. The proposed device uses a clocking scheme to ensure non-reciprocity by initializing the receiving magnet at the saddle point of the energy landscape. Moreover, the clocking scheme enables the device to be cascaded in a domino logic scheme; thus, the overall delay and the energy dissipation of a more complicated circuit like a 32-bit ALU further improves. In this paper, the impact of pulse skew and amplitude on the delay and the energy of the proposed device is studied as well.

The rest of the paper is organized as follows. Section II elaborates on the operation and the modeling of the proposed magnetostriction-assisted all-spin logic (MA-ASL) device and clocking circuitry. Section III benchmarks the energy dissipation and the delay of a 32-bit MA-ASL ALU against selected CMOS and the original ASL. Section IV analyzes the impact of thermal noise and pulse skew on the operation and the performance of the MA-ASL device. Section V provides the conclusion.

This work was funded by Semiconductor Research Corporation (SRC) under the research program MSR-Intel with research task/theme ID spin interconnects-2616.001.

R. Mousavi Iraei, N. Kani, S. Dutta, and A. Naeemi are with the Department of Electrical and Computer Engineering, Georgia Institute of Technology, Atlanta, GA, USA (e-mail: iraei@gatech.edu).

D. E. Nikonov, S. Manipatruni, and I. A. Young are with Components Research Group, Intel Corporation, Hillsboro, OR, USA.

J. T. Heron is with the Department of Materials Science and Engineering, University of Michigan, Ann Arbor, MI, USA.

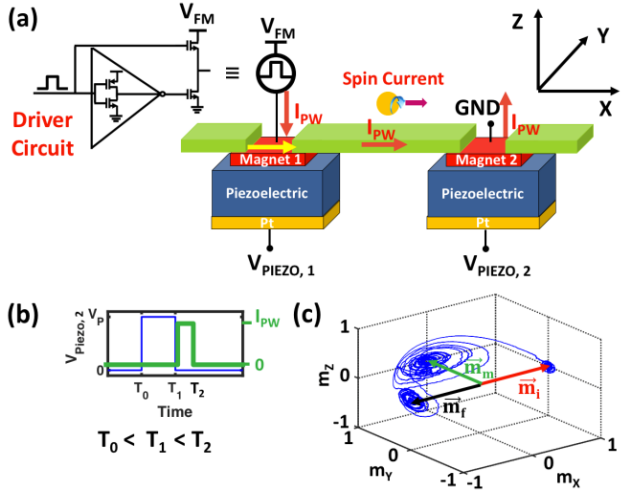

Fig. 1: (a) Schematics of the proposed device and driver circuit. (b) By applying the pulse, $V_{PIEZO,2}$, at $T_0$, Magnet 2 reorients from a stable state the +x direction to the meta-stable state, the y direction. Then, $V_{PIEZO,2}$ is turned off, and the current provided by the driver circuit, $I_{PW}$, is turned on. The current becomes spin polarized when passing through Magnet 1, oriented in the +x direction, and applies a torque to Magnet 2, which reorients from the y direction to the −x direction, the other stable state. (c) First, the orientation of Magnet 2, rotates by 90⁰ from $\vec{m}_i$ to $\vec{m}_m$ using magnetostrictive switching; then, it reorients by 90⁰ from $\vec{m}_m$ to $\vec{m}_f$ using the spin-transfer torque.

## II. DEVICE OPERATION

A single stage MA-ASL device is composed of a transmitter Magnet 1 and a receiver Magnet 2 connected via a metallic (Cu) channel forming a CSV structure, Fig. 1a. First, the receiver magnet is reoriented due to magnetostrictive switching due to the voltage applied to the piezoelectric layer at the receiver side, $V_{PIEZO, 2}$, from $T_0$ to $T_1$, as shown in Fig. 1b. The applied voltage generates an anisotropic strain that couples to Magnet 2 altering the magnetoelastic energy of the magnet; therefore, the easy axis rotates from the x to the y direction. As a result, Magnet 2 rotates to the y direction, Fig. 1c. By switching off $V_{PIEZO,2}$ at $T_1$, the easy axis rotates back to the x direction; thus, the magnet will be placed at the meta-stable saddle point of its energy profile, and the magnet will be equally probable to fall into the +x or −x directions. To break the symmetry, an electrical current with a spin polarization opposite to the orientation of Magnet 1 is applied to Magnet 2 from $T_1$ to $T_2$, forcing Magnet 2 to rotate to the -x direction, opposite to the orientation of Magnet 1, as shown in Fig. 1a, Fig. 1c. Therefore, the proposed structure acts as an inverter. Providing a large pulse current at a very low voltage would require a large portion of the energy to be dissipated in the driver transistors. Hence, the driver circuit, shown in Fig. 1a, is proposed to more efficiently generate the required pulse current. To reduce the energy dissipation, the voltage drop on the transistor is limited to the $V_{DS}$ voltage of a CMOS transistor.

To model the operation of driver transistors, predictive technology models [26] are utilized, while SPICE models are developed to model the operation of spintronic parts following a similar approach taken in [27]. To account for the physics of the device, we need to self-consistently solve the equations governing the dynamics of the magnetization, spin transport in the metallic channel, and magnetostriction. The magnetization dynamics is explained by the stochastic LLG equation

$$\frac{d\vec{m}}{dt} = -\gamma\mu_0\left[\vec{m} \times \vec{H}_{eff}\right] + \alpha\left[\vec{m} \times \frac{d\vec{m}}{dt}\right] + \frac{\vec{I}_{s,\perp}}{qN_s}, \quad (1)$$

in which $\vec{m}$, $\vec{I}_{s,\perp}$, $N_s$, $\mu_0$, $\alpha$, $\gamma$ represent the magnetic orientation, the perpendicular spin current, the number of spins in the magnet, the free space permeability, the Gilbert damping coefficient, and the gyromagnetic ratio [27] - [28]. The net magnetic field, $\vec{H}_{eff}$, is comprised of the uniaxial anisotropy field, $\vec{H}_U$, the demagnetization field, $\vec{H}_{demag}$, and the thermal noise field, $\vec{H}_{Thermal}$, which models the statistical thermal motion of the electrons [27]. For the MA-ASL device, the anisotropic field, $\vec{H}_U$, is due to the variations in the magnetoelastic energy, $E_{ME}$ [29],

$$E_{ME} = -\frac{3}{2}\lambda Y\left[\left(m_x^2 - \frac{1}{3}\right)\epsilon_{xx} + \left(m_y^2 - \frac{1}{3}\right)\epsilon_{yy} + \left(m_z^2 - \frac{1}{3}\right)\epsilon_{zz}\right]. \quad (2)$$

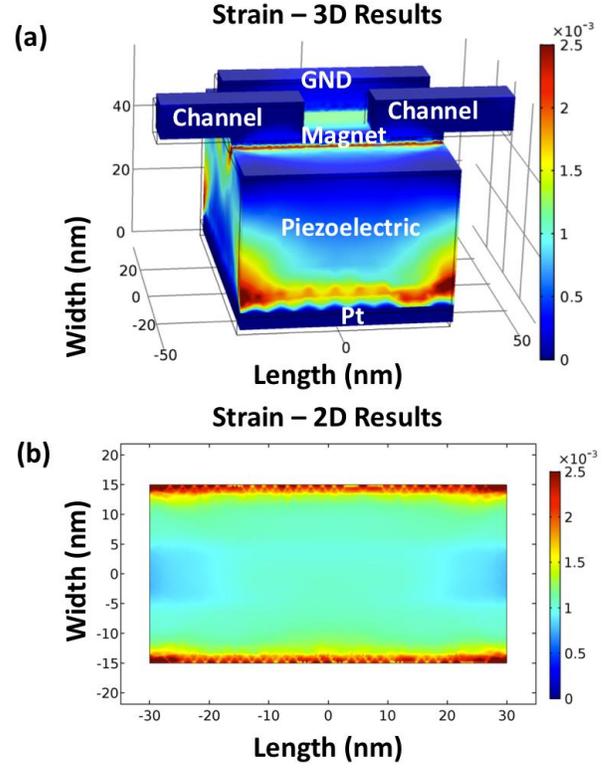

Fig. 2: Generated strain is simulated using a Comsol model developed from the piezoelectric parameters of PMN-PT [30]. Results shown for (a) the device and (b) the cross-section of the magnet demonstrate that the net strain of 1200 ppm will be transferred to the magnet.

In (2), $\lambda$ and $Y$ represent the magnetostrictive coefficient and the Young's modulus, respectively, and $m_x$, $m_y$, and $m_z$ are the magnetic orientation along the x, the y, and the z directions, respectively. In (2), $\epsilon_{xx}$, $\epsilon_{yy}$, and $\epsilon_{zz}$ are components of the strain matrix; hence, the anisotropic field is derived as

$$\vec{H}_U = -\frac{1}{\mu_0 M_S}\frac{\partial E}{\partial \vec{m}} = \frac{2K}{\mu_0 M_S}m_Y\vec{Y}, \quad (3)$$

where $\mu_0$ and $M_S$ represent the permeability of free space and saturation magnetization. In (3), the energy density, $K$, due to the magnetostriction, is proportional to the net anisotropic strain [23],

$$K = \frac{3}{2}\lambda(C_{11} - C_{12})(\epsilon_{yy} - \epsilon_{xx}), \quad (4)$$

where $C_{11}$ and $C_{22}$ are elastic stiffness constants [23],

$$\epsilon_{xx} = \epsilon_0 + d_{31}\frac{V_{PIEZO}}{t}, \quad (5)$$

$$\epsilon_{yy} = \epsilon_0 + d_{32}\frac{V_{PIEZO}}{t}, \quad (6)$$

in which $d_{31}$ and $d_{32}$ are piezoelectric constants. The transferred anisotropic strain to the magnet is investigated using COMSOL based on PMN-PT material parameters (Table 1) [30]. As Fig. 2 shows, a large net anisotropic strain $(\epsilon_{yy} - \epsilon_{xx})$ of 1200 ppm is transferred to the magnet, when $V_{PIEZO,2}$ is 100 mV. The generated strain is large enough to reorient magnets by $90^0$.

Table 1
Simulation Parameters

| Piezoelectric (PMN-PT) [30] | | |
|---|---|---|
| Piezoelectric Constant | $d_{31}$ | 813 pC/N |
| Piezoelectric Constant | $d_{32}$ | -2116 pC/N |
| Piezoelectric Height | t | 40 nm |
| Magnets (Terfenol-D) [31] - [33] | | |
| Magnet Length | $L_X$ | 60 nm |
| Magnet Width | $L_Y$ | 30 nm |
| Magnet Height | $L_Z$ | 3 nm |
| Saturation Magnetization | $M_S$ | 1.0 T |
| Magnet Barrier | $\Delta/K_B T$ | 40 |
| Damping Factor | $\alpha$ | 0.1 |
| Spin Polarization | P | 0.8 |
| Resistivity | $\rho$ | $60 \times 10^{-8} \Omega.m$ |
| Magnetostrictive Coefficient | $\lambda$ | 1200 ppm |
| Young's Modulus | $\gamma$ | 90 GPa |
| Channels (Cu) [8], [34] | | |
| Channel Length | $L_{Int}$ | 132 nm |
| Channel Width | $W_{Int}$ | 44 nm |
| Aspect Ratio | AR | 1 |
| Conductivity | $\sigma$ | 27.4 1/$\mu\Omega m^2$ |
| Grain Boundary Reflection Probability | R | 0.2 |
| Specularity Parameter | P | 0.0 |
| Spin Relaxation Time | $\tau_S$ | 8.92 ps |
| Electron Mobility | $\mu$ | 0.002 $m^2$/Vs |

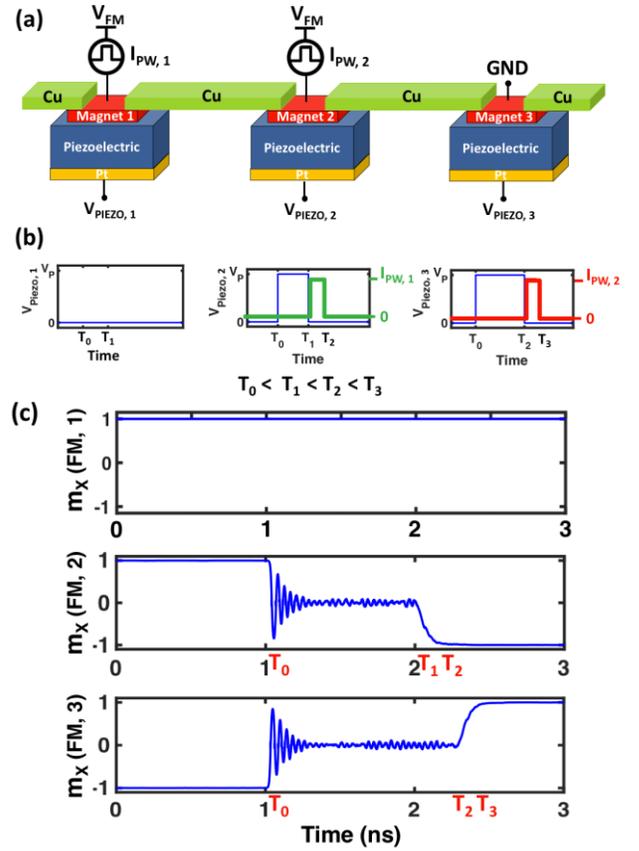

Fig. 3: (a) Three magnets cascaded in a domino logic scheme. (b) Piezoelectric pulses are applied simultaneously at $T_0$ and released sequentially at $T_1$ and $T_2$ to perform the first $90^0$ of the switching. The current pulses, provided by the clocking circuit, are applied at $T_1$ and $T_2$ to perform the second $90^0$ of the switching via STT. (c) The overall delay of the inverter chain significantly reduces by simultaneously performing the first $90^0$ of the switching for all magnets of a chain.

Spin transport in the metallic channel, connecting the input magnet to the output magnet, is accounted using the drift-diffusion equation for the overall spin density,

$$\frac{\partial s}{\partial t} = D\frac{\partial^2 s}{\partial x^2} - \mu E\frac{\partial s}{\partial x} - \frac{s}{\tau_s}, \quad (7)$$

in which $s$, $D$, $\mu$, $E$, and $\tau_s$ are the spin density, the diffusion coefficient, the electron mobility, the electric field intensity, and the spin relaxation time, respectively [27]. The equation is modeled using the circuit elements, presented in [27] for a metallic channel, calibrated with experimental results [34]. The impact of variations of spin relaxation mechanism, due to size effects and geometrical dimensions, on the delay and the energy dissipation of a metallic channel is studied in [12]. Spin transport equation for the metallic channel, magnetostriction, and stochastic LLG equations are solved self-consistently using SPICE simulations. The simulation parameters are shown in Table 1. Moreover, simulations are done for a 3-stage cascaded inverter chain of magnets as illustrated in Fig. 3a. First, Magnet 2 and Magnet 3 are initialized in the y direction by applying piezoelectric voltage pulses, Fig. 3b. Then, voltage pulses are turned off sequentially, and current pulses are applied to perform the second $90^0$ switching. As shown in Fig. 3c, by initializing the magnets simultaneously, the first $90^0$ of magnetization switching, which takes ~1 ns, is more than 10 times larger than the second $90^0$ switching, but is shared between the two magnets; thus, the overall delay is improved. The benefit of this approach obviously grows as the logic depth (the number of cascaded gates) increases.

## III. 32-BIT MA-ASL ALU

At the heart of an arithmetic logic unit (ALU) is the arithmetic operations (AO) block, which performs operations such as addition, subtraction, NAND, and NOR. For a 32-bit ALU, operations are done on two 32-bit input numbers A and B. In contrast to NAND and NOR operations, addition and subtraction cannot be done in parallel. As Fig. 4 illustrates, the addition of A and B is done by a 32-bit ripple carry adder in which the result is S. The addition operation requires propagating the carry signal, $C_i$ bits, in the critical path from one bit to the next bit. Therefore, the propagation delay of carry bits across the 32-bit adder dominates the delay of an ALU. For a 32-bit MA-ASL ripple-carry adder, the critical path is comprised of 32 full adders each consisting of two magnets in the path. In a cascade scheme like Fig. 3, all 64 magnets are initialized simultaneously; thus, the overall delay is

$$\tau_{32-Bit-Adder} = \tau_{ME} + 64\tau_{STT}, \qquad (8)$$

where $\tau_{32-Bit-Adder}$, $\tau_{ME}$, and $\tau_{STT}$ are the delay of the 32-bit adder, the initializing time of 1 ns due to magnetostrictive switching, and the delay of switching from the saddle point due to STT, about 35 ps for an error rate below $10^{-3}$, respectively. Thus, the overall delay of the 32-bit adder will be 3.3 ns. By accounting for the 32-bit adder, repeaters at each 300 nm, NAND and NOR gates, and other gates of a 32-bit MA-ASL ALU, the delay and the energy dissipation will be 11.8 ns and 5.2 pJ, respectively. Compared to the delay and the energy dissipation of a 32-bit ASL ALU, those of MA-ASL show 21x and 27x improvement, respectively, Fig. 5. However, the delay and the energy of the device compared to those of TFETs and CMOS devices are still larger. Although spintronic devices cannot compete against CMOS devices in Boolean applications, such as 32-bit adders and ALUs, these devices may compete against CMOS in non-Boolean applications because of efficient implementation of majority gates in spin-based devices. Furthermore, by accounting for the significant improvement of the energy and the delay of the MA-ASL compared to those of ASL, the device may become competitive against CMOS for non-Boolean applications. Even in the case of Boolean computations, by taking advantage of pipelining in the design of complicated systems such as 32-bit ALUs, slow and low-energy devices may become more competitive. In a pipelining scheme, the output magnet of $FA_{i+1}$ will be initialized right after the $C_i$ bit is generated by $FA_i$. Thus, $FA_i$ can immediately operate on the next bit in line without waiting for the previous bit to propagate to the last magnet in the line, which represents $C_{32}$.

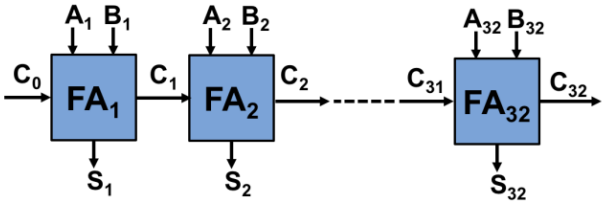

Fig. 4: 32-bit ripple carry adder consisted of 32 full adders (FAs). Carry bits propagate through the critical path, comprised of 64 magnets.

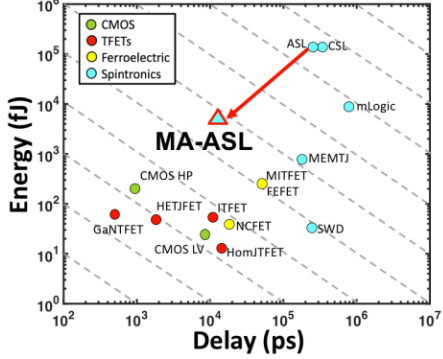

Fig. 5: Energy dissipation and delay of various spintronic, CMOS, and TFET technologies for implementation of a 32-bit ALU. The delay and the energy dissipation of MA-ASL ALU compared to those of ASL ALU, show 21x and 27x improvement, respectively. [20].

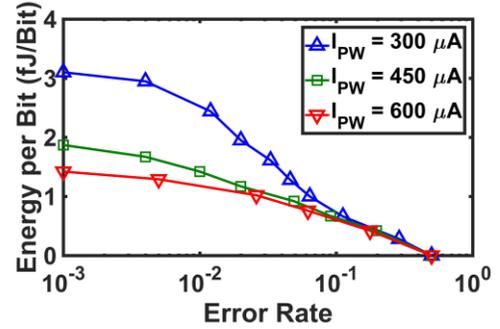

Fig. 6: Increasing the amplitude of the pulse current, lowers the required pulse width to reach to the same error rate; thus, as pulse amplitude increases, lower energy is required to reach to the same error rate. However, the amplitude of the pulse current cannot exceed certain maximum limits due to electromigration [35].

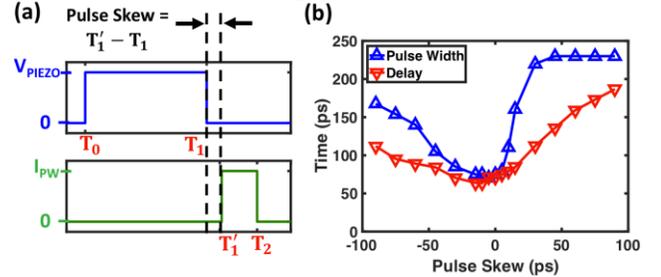

Fig. 7: (a) Definition of pulse skew. (b) The required pulse width and delay to reach to an error rate of $10^{-3}$. Positive compared to negative pulse skew, has more impact on increasing delay; hence, the device can be designed with a small embedded negative pulse skew to offset any undesired positive pulse skew due to fabrication limitations and inaccuracies.

In this scheme, the delay to generate the last bit, $C_{32}$, is $32\tau_{ME} + 64\tau_{STT}$, larger than the delay of the domino MA-ASL adder, explained above. However, a new result is generated each $\tau_{ME} + 2\tau_{STT}$ instead of each $\tau_{ME} + 64\tau_{STT}$. Thus, the throughput of the 32-bit adder, the heart of an ALU block, further increases.

## IV. Clocked MA-ASL

Operating the MA-ASL device either as a domino logic or as a pipeline, requires a clocking scheme that precisely accounts for the times required to perform the first half of the switching, done through magnetostrictive switching, and the second half of the switching, done through applying STT. The $90^0$ magnetic rotation time under STT is inversely proportional to the amplitude of the pulse current [28]; thus, by increasing the amplitude of the pulse current, delay decreases; hence, the width of the required pulse current decreases; thus, lower energy must be dissipated to reach to a certain error rate, Fig. 6. However, the amplitude of the pulse current is limited to the maximum current, not reaching to electromigration [35]. The energy dissipation is mainly associated with three parts: (1) the transistors of the driver circuit, illustrated in Fig. 1a, about 2 fJ, (2) ohmic energy dissipation inside the MA-ASL device during the STT switching, about 0.2 fJ, and (3) in the form of $CV^2$ to provide pulse voltages of the piezoelectric, in the range of a few aJs. We have accounted for these factors in calculating the total energy dissipation of the device.

The proper operation of the proposed domino logic and the pipelining schemes, depends on turning off the piezoelectric pulses and applying spin current pulses simultaneously at $T_1$, Fig. 1b. However, due to factors including the inaccuracies and the limitations of the fabrication processes, the two clocks perform with a negative or a positive pulse skew defined as $T_1' - T_1$ in Fig. 7a. At $T_1$, Magnet 2 is placed at the meta-stable saddle point of the energy profile and equally probable to fall into the stable directions, +x or –x. In the absence of STT due to a positive pulse skew, the magnet starts to randomly switch to one of the stable directions; hence, when STT is applied at $T_1'$, magnetization is deviated from the **y** axis, the meta-stable point of energy, by an angle $\theta'_m$; thus, the longer pulse width and switching time are expected (Fig. 7b). Increasing the positive pulse skew, increases the angle, $\theta'_m$; hence, delay further increases. Although a negative pulse skew may contribute to a non-zero $\theta'_m$, the deviations in the case of a negative pulse skew compared to that of a positive pulse skew will be smaller due to the presence of the magnetoelastic energy. Thus, designing an MA-ASL-based circuit with a small embedded negative pulse skew about 5-10 ps offsets probable undesired positive skews due to fabrication inconsistencies.

## V. Conclusion

Studies have examined ASL devices for various Boolean and non-Boolean applications owning to their efficient implementation of majority gates, low voltage operation, and non-volatile memory. This paper proposes an ASL-based heterostructure of magnets and piezoelectrics that employs both magnetostriction and STT to perform magnetization reversal. The proposed device excels in domino logic and pipelining schemes using the driver circuit, proposed in the paper. The performance of the device is benchmarked against ASL, TFETs, and CMOS technologies. The paper illustrates that the energy and the delay performance of a 32-bit ALU designed by the MA-ASL device compared to those of the ASL device show 21x and 27x improvement, respectively. However, the device cannot compete against CMOS devices in implementing Boolean functions, but the device, augmented by the advances in piezoelectric and magnetic materials, may become competitive against CMOS in implementing non-Boolean functions.

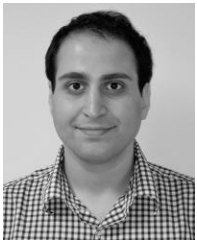

**Rouhollah Mousavi Iraei** (S'11) received a B.S. degree in electrical engineering from Sharif University, Tehran, Iran, in 2012 and an M.S. degree in electrical and computer engineering from the Georgia Institute of Technology in 2015. He is currently pursuing a Ph.D. degree in electrical and computer engineering with the Georgia Institute of Technology, Atlanta, GA, USA.

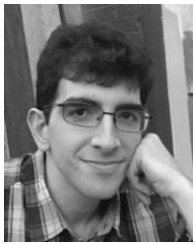

**Nickvash Kani** received his B.S. degree in Computer Engineering from Boston University, Boston, MA, in 2010 and M.S. and Ph.D. degrees in Electrical and Computer Engineering from the Georgia Institute of Technology, Atlanta, GA in 2013 and 2017, respectively.

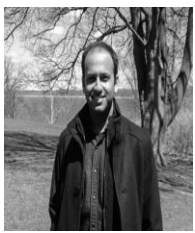

**Sourav Dutta** (S'13) received the B.E. degree in Electrical Engineering from Jadavpur University, Kolkata, India, in 2012. He is currently pursuing his Ph.D. degree in Electrical and Computer Engineering at Georgia Institute of Technology, Atlanta, GA, USA under Dr. Azad Naeemi. His primary area of researc is modeling and simulation of Spintronic devices and interconnects for Beyond-CMOS application with special focus on spin waves which is funded by and in collaboration with Intel. From May to July 2016, he was a visiting researcher at IMEC, Belgium where he was involved in modeling and simulation of nanoscale plasmonic logic gates, magneto-plasmonics and spin-plasmonics for boolean and non-boolean computation for Beyond-CMOS application.

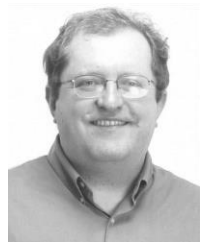

**Dmitri E. Nikonov** (M'99–SM'06) received an M.S. degree in aeromechanical engineering from the Moscow Institute of Physics and Technology, Moscow, Russia, in 1992, and a Ph.D. degree in physics from Texas A&M University, College Station, TX, USA, in 1996. He is currently a Principal Engineer with the Components Research Group, Hillsboro, OR, USA, where he is involved in simulation and benchmarking of beyond CMOS logic devices.

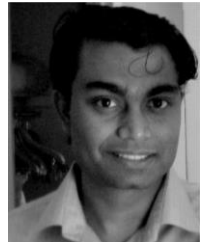

**Sasikanth Manipatruni** (M'07) received a B.S. degree in electrical engineering from IIT Delhi, Delhi, India, and a Ph.D. degree in electrical and computer engineering from Cornell University, Ithaca, NY, USA. He is a Staff Scientist with the Exploratory Integrated Circuits Group, Components Research Intel, Hillsboro, OR, USA, where he is involved in beyond-CMOS devices and circuits.

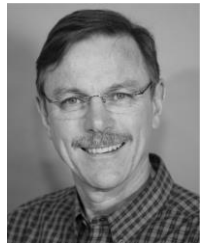

**Ian A. Young** (M'78–SM'96–F'99) received a B.S. degree in electrical engineering and an M.S.E. degree from the University of Melbourne, Melbourne, VIC, Australia, and a Ph.D. degree in electrical engineering and computer science from the University of California at Berkeley, Berkeley, CA, USA. He is a Senior Fellow with the Technology and Manufacturing Group, Intel Corporation, Santa Clara, CA, USA.

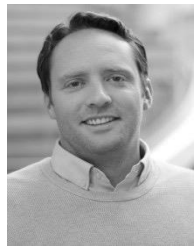

**John T. Heron** studied Physics as an undergraduate at the University of California, Santa Barbara. As a graduate student, he was awarded the NDSEG graduate fellowship and worked in the thin film complex oxide group of Professor Ramamoorthy Ramesh at the University of California, Berkeley. In 2013 John was awarded Ross N. Tucker Memorial Award. After earning his masters (2011) and doctoral (2013) degrees from the University of California Berkeley he began postdoctoral research at Cornell University under the co-mentorship of Professors Darrell Schlom and Dan Ralph. John joined the Department of Materials Science and Engineering as an Assistant Professor in Winter 2016.

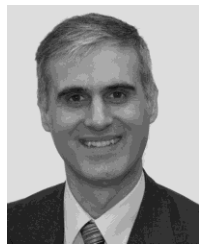

**Azad Naeemi** (S'99–M'04–SM'04) received a B.S. degree in electrical engineering from Sharif University, Tehran, Iran, in 1994, and M.S. and Ph.D. degrees in electrical and computer engineering from the Georgia Institute of Technology (Georgia Tech), Atlanta, GA, USA, in 2001 and 2003, respectively. He is currently an Associate Professor with the School of Electrical and Computer Engineering at Georgia Tech.